\documentstyle[referee]{aa}
\begin{document}

\title{ Basic properties of Gamma-ray 
loud blazars}

\author{K.S. Cheng\inst{1} \and J.H. Fan\inst{2} \and L. Zhang\inst{1,3}}
\institute{Department of Physics, The University of Hong Kong,
Pokfulam Road, Hong Kong, China \and Center for Astrophysics, 
Guangzhou Normal University, Guangzhou 510400, China \and 
 Department of Physics, Yunnan University, Kunming, China}

\maketitle
\begin{abstract}
In this paper, a method is proposed to determine the basic properties
 of $\gamma$-ray loud blazars, among them  the central 
 black hole mass, $M$, the Doppler factor, $\delta$, the propagation
 angle of the $\gamma$-rays with respect to the symmetric axis of
 a two-temperature  accretion disk, $\Phi$, 
 and the distance (i.e. the height above the accretion
 disk), $d$  at which the $\gamma$-rays are created,
 for seven $\gamma$-ray loud blazars with available GeV variability
timescales and in which  the absorption  effect of a $\gamma$-ray and the
beaming effect have been taken into account. Our results indicate 
that, if we take the intrinsic $\gamma$-ray luminosity to be  $\lambda$ 
times the Eddington luminosity, $L_{\gamma}^{in} = \lambda L_{Edd.}$,
 the masses of the blazars are in the 
 range of $(4 \sim 83)\times 10^{7}M_{\odot}$ ($\lambda = 1.0$) or $(6
\sim 131)\times 10^{7}M_{\odot}$ ($\lambda = 0.1$), the
 Doppler factors ($\delta$) lie in the range of 0.57 to 3.72 
 ($\lambda = 1.0$) or 0.79 to 5.33
($\lambda = 0.1$), the angle ($\Phi$) is in the range of
 $14^{\circ}$ to 43$^{\circ}$ ($\lambda = 1.0$) or
 $13^{\circ}$ to 39$^{\circ}$ ($\lambda = 0.1$), and the distance  ($d$)
 is in the range of 
 28$R_{g}$ to 411$R_{g}$ ($\lambda = 1.0$) or 26$R_{g}$ to 366$R_{g}$
($\lambda = 0.1$). 
 For 3C279, the case of a uniformly-bright disk is also adopted to
 determine the basic parameters, which are compared with those 
 obtained for a two-temperature disk as well as  those obtained 
 by Becker \& Kafatos (1995).
 Our model results are independent of $\gamma$-ray emission mechanisms but
 they do depend on the X-ray emission mechanism of the accretion disk.
 
\end{abstract}


\section{Introduction}

 One of the most important results of the CGRO/EGRET instrument in 
 the field of 
 extragalactic astronomy  has been  the discovery of  blazars (i.e.
flat-spectrum 
 radio quasars--(FSRQs) and BL Lac objects)  which emit most of their
bolometric 
 luminosity in the high energy range of $\gamma$-rays ( E $>$ 100 MeV ).  
 Many of the $\gamma$-ray emitters are also superluminal radio sources 
 (von Montigny et al. 1995).  The common properties of these EGRET-detected 
 AGNs are the  following:  the $\gamma$-ray flux is dominant over the flux 
 in lower energy bands; the $\gamma$-ray luminosity above 100 MeV ranges 
 from $\sim$ $3 \times 10^{44}$ erg s$^{-1}$  to $\sim$ 
  $10^{49}$ erg s$^{-1}$ 
 (assuming isotropic emission); many of the sources are strongly variable
in 
 the $\gamma$-ray band on timescales varing from days to months
(Mukherjee et al. 
 1997), but large flux  variability on short timescales of $< 1$ day is  
 also detected ( see below ).  These facts suggest that the $\gamma$-ray
emission is  
 likely to arise from the jet of a blazar. 

 Various  models for $\gamma$-ray emission from AGNs have been proposed.
 Generally, they are of two kinds: leptonic and hadronic models.
 In the leptonic model, high energy $\gamma$-rays are produced by the
inverse Compton scattering of  high energy electrons in the soft
 photon field. The soft photons may be emitted from the 
 nearby accretion disk 
 ( Dermer et al.  1992; Coppi et al. 1993; Zhang \& Cheng, 1997a)  or
 they may arise from disk radiation 
 reprocessed  in some region of AGNs ( e.g. a broad emission line region)  
 (Sikora et al.  1994; Blandford \& Levinson 1995; Xie et al. 1997, 1998);
or they may come from
the  
 synchrotron emission in the jet ( Maraschi et al. 1992;  Zdziarski 
 \& Krolik 1993; Bloom \& Marscher 1996; Marscher \& Travis 1996), or
 from a differential rotating flux tube near the inner edge of the
 accretion disk (Cheng, Yu \& Ding 1993).
 In the hadronic model, high energy $\gamma$-rays are produced by the  
 synchrotron emission from ultrarelativistic electrons and positrons 
 created in a  proton-induced cascade ($PIC$) (Mannheim \& Biermann 1992; 
 Mannheim 1993; Cheng \& Ding 1994). 
 TeV radiation has been observed from 4 X-ray-selected BL Lacertae
objects (XBLs):
 Mkn 421 ( Punch et al. 1992), Mkn 501 (Quinn et al. 1996) and 
 IES 2344+514 (Catanese et al. 1998), PKS 2155-304 (Chadwick et al. 1999).
 But there is no consensus  yet on the dominant emission process (see 
 von Montigny et al. (1997) for 3C273, 
 Ghisellini et al. (1996) for 3C279, Comastri et al. (1997) for 0836+710, 
B$\ddot{o}$ttcher \& Collmar 1998 for PKS 0528+134).

It is generally believed that the escape of high energy $\gamma$-rays from
an AGN depends on $\gamma-\gamma$ pair production process because there
are lots of soft photons around the central black hole. Becker \& Kafatos
(1995) have calculated the $\gamma$-ray optical depth in the X-ray field
 of an accretion disk. They found that the $\gamma$-rays should escape
preferentially along the symmetric axis of the disk, due to the strong
 angular dependence of the pair production cross section. The phenomenon
 of $\gamma-\gamma$ "focusing" is related to the more general issue of
 $\gamma-\gamma$ transparency, which sets a minimum distance between the
 central black hole and the site of $\gamma$-ray production (
 Bednarek 1993, Dermer \& Schlickeiser 1994, Becker \& Kafatos 1995,
 Romero et al. 2000; Zhang \& Cheng 1997b). Therefore the $\gamma$-rays
are focused in a small
 solid angle, $\Omega = 2\pi(1-cos \Phi)$, suggesting that the apparent 
  observed
 luminosity should be expressed as $L_{\gamma}= \Omega
D^{2}(1+z)^{\alpha_{\gamma}-1}F^{obs.}_{\gamma}(>100MeV)$, where
$F_{\gamma}^{obs.}$ is the
observed energy flux of the $\gamma$-rays, $D$ the distance to the AGN,
and
$z$ the redshift.
The observed $\gamma$-rays from
the AGN
require that the jet almost points to us and the optical depth $\tau \leq
1.0$. In this sense, both the absorption and beaming ( boosting ) effects
should be considered when the properties of a $\gamma$-ray loud blazar are
discussed, which is the focus of the present paper. The paper is arranged as 
 follows: In section 2 we summarize the observed results of seven
$\gamma$-ray loud blazars with available GeV variable timescale. In
section 3, we present our method which is used to estimate the black
 hole mass, the Doppler factor, the propagation angle of the
$\gamma$-rays, and their emission distance above the accretion disk.
  In section 4, we discuss the results and give a brief summary.

H$_{0}$ = 75 km s$^{-1}$ Mpc$^{-1}$, and q$_{0}$ = 0.5 are adopted 
throughout the paper.

\section{Observed data}

 Since we are interested in the variability timescale, we consider
 here only those $\gamma$-ray loud blazars with short
 timescales of variation, detected in the $\gamma$-ray region. 
 Because the variability timescale corresponds to a different
 variation amplitude for different sources and/or different observation
 periods, we use the doubling timescale,  
 $\Delta T_{D} = (F_{minimum}/\Delta F) \Delta T$, as the variability
 timescale, where $\Delta F = F_{maximum} - F_{minimum}$ is the variation
 of the flux over the time  $\Delta T$.
There are few simultaneous observations of the X-ray and
 $\gamma$-ray bands available, the data considered here are not
simultaneous. The
 $\gamma$-ray data are from those periods corresponding to variability. 
 The X-ray
 data are from recent publications, particularly the paper by
 Comastri et al. (1997).

\subsection{PKS0208-512}
 PKS0208-512 is an HPQ ($P_{opt} = 11.3\%$, see Impey and Tapia, 1988). It
 was visible to EGRET during the 1991 (Bertsch et al. 1993) and 1994/1995
 (Stacy et al. 1996) observation periods.  The 1991 Sep./Oct. observations
 can be expressed well by a power law with a photon spectral index of
 $\alpha_{ph} = 1.689$ and an isotropic luminosity of
 $L{(>100MeV)} = 2~10^{48} erg~s^{-1}$.  The 1991 observations show
evidence for a variability of at least a factor of 3 on a time scale of
tens of days. During the 1994/1995 observation period a variation of a
factor of 2.5 over a time scale of a two-week EGRET point was detected
(Stacy et al. 1996) suggesting $\Delta T_{D} = 5.6~ days$.  $f_{1KeV} =
0.61 \mu $Jy and $\alpha_{X} = 1.04$ are available in the paper by
 Comastri et al. (1997).
    
 \subsection{PKS 0528+134}

 PKS 0528+134, $z = 2.07$ (Hunter et al. 1993), is one of the most 
 luminous examples of blazars. It is observed by EGRET, COMPTE and OSSE aboard
 the CGRO (see Hunter et al. 1993,  McNaron-Brown et al. 1995, 
 Mukherjee et al. 1996, Collmar et al. 1997; Sambruna et al.  1997).

 During 23-29 March 1993, 
   F($>$100 MeV) = (0.23$\pm$0.12 -- 3.08 $\pm$ 0.35) 
 $\times 10^{-6}$ photon cm$^{-2}$ s$^{-1}$, with a photon spectral 
 index $\alpha_{ph}$ = 2.21$\pm$0.10.  In the 1993 observation, 
 a variation of order of 100\% over a timescale of $\sim$
 2 days was detected (see Wagner et al. 1997), which suggests a doubling
time
 scale of $\Delta T_{D}$ = 1day. X-ray data of $f_{1KeV} = 0.65
 \mu$ Jy and $\alpha_{X} = 0.54 \pm 0.29$ are reported in the paper by
 Comastri et al. (1997).

\subsection{PKS 0537-441}
 
 PKS 0537-441, $z$ = 0.896, a candidate for a gravitational lens 
 (Surpi et al. 1996 ), is a violently variable object (Romero et al. 1995; 
Fan \& Lin 1999). 
 The  $\gamma$-ray flux varies from (1.83$\pm$0.91) to 
 (8.98$\pm$1.45) $\times 10^{-6}$ photon cm$^{-2}$ s$^{-1}$ 
 (Mukherjee et al. 1997). A flare of a factor of $\sim$3 
 from 0.35 to 2.0 $\times 10^{-6}$ photon cm$^{-2}$s$^{-1}$ over a time 
 scale of $\sim$ 2 days can be seen from Fig. 3 in  Hartman's paper 
 (Hartman 1996), $\Delta T_{D}$ = 16 hrs. $f_{1KeV} = 0.81 \mu$ Jy and
 $\alpha_{X} = 1.16 \pm 0.09$ (Comastri et al. 1997).

\subsection{1253-055~ (3C279)}

 3C279 is a well known member  of the OVV subclass of blazars.  It is
perhaps 
 the prototypical superluminal radio source (Moffet et al. 1972) and the 
 first quasar detected at energies greater than 1 GeV by EGRET/CGRO.  
 The simultaneous variability in X-rays and $\gamma$-rays ($ > $ 100
MeV) 
 suggests for the first time that they are approximately  cospatial (
 M$^{c}$Hardy 1996). The $\gamma$-ray flux varies from 1.28 to 28.7 
 $\times 10^{-6}$ photon cm$^{-2}$ s$^{-1}$ (Mukherjee et al. 1997). 
 Two $\gamma$-ray flares were detected (see Kniffen et al. 1993; 
 Hartman et al. 1996; M$^{c}$Hardy 1996; Wehrle et al. 1998).

 The 16-28 June 1991 flare showed: 
  F($>$100 MeV) = ( 2.8 $\pm$ 0.4) $\times 10^{-6}$ photon cm$^{-2}$ s$^{-1}$
  with a photon spectral index $\alpha_{ph}$ = 1.89$\pm$0.06. A variation 
 of a factor of 4 over 2 days was seen. The quasi-simultaneous X-ray
 data detected in June 1991  can be described by $J(E) = (4.0 \pm 0.34)
\times 10^{-3} E^{-(1.68 \pm 0.05)}$ photon~cm$^{-2}$~s$^{-1}$ Kev$^{-1}$
(Hartman et al. 1996).

 The January-February 1996 flare intensity was ( see McHardy 1996; Wehrle
et al. 
 1998), 
  F($>$100 MeV) = ( 11.0 $\pm$ 1.) $\times 10^{-6}$ photon cm$^{-2}$ s$^{-1}$
  with a photon spectral index of $\alpha_{ph}$ = 1.97$\pm$0.07. During
this
 flare, a variation 
 of a factor of 4$\sim$5 in a day was observed,  
 $\Delta T$ = 6 hrs (Wehrle et al. 1998). During the 1996 January
observation
 period, 3C279 was detected at a level of $3\times10^{-3}$
photon~cm$^{-2}$~s$^{-1}$KeV$^{-1}$ at 1 KeV with a spectral index 
$\alpha_{X}=0.78$ (Lawson \& M$^{c}$Hardy, 1998).
 
\subsection{PKS 1622-297}
 
 For PKS 1622-297, $z$=0.815, we have very little information in the lower 
 energy bands. But it
 is one of the most luminous objects in the $\gamma$-ray region. A peak 
 flux of (17$\pm$3)$\times 10^{-6}$ photon cm$^{-2}$ s$^{-1}$ (E$>$100 MeV) 
 and a
 flux increase by a factor of 2 in 9.7 hours  were observed ( Mattox et al. 
  1997).  $f_{1KeV} = 0.08 \mu$ Jy  ( Mattox et al. 1997).

\subsection{Q1633+382~(4C 38.41)}
 
 Quasar 1633+382, $z=1.814$, is an LPQ (P$_{opt}=2.6\%$, Moore \& Stockman 
 1984). During the 1992 November 17 - December 1 observation period, it
was
detected with a flux of 
 F($>$100 MeV) = ( 0.30 $\pm$ 0.06) $\times 10^{-6}$ photon cm$^{-2}$
s$^{-1}$,
  with a photon spectral index $\alpha_{ph}$ =  1.87$\pm$0.07. 
 The flux varied by a factor of 1.5 within 24 hrs, $\Delta T_{D}$ = 16
hrs,
  while the  spectral index did not change. The $\gamma$-ray luminosity is 
 at least  two orders of magnitude larger than the maximum ever observed 
 in any  other band (see Mattox et al. 1993).  $f_{1KeV} = 0.42 \mu$ Jy
and
 $\alpha_{X} = 0.53 \pm 0.08$ (Comastri et al. 1997).

\subsection{2200+420~(BL Lacertae)}
 
 2200+420 is the prototype of the BL Lacertae class. It is  variable at
all
 wavelengths (see Fan et al. 1998a, 1998b; Bloom et al. 1997;
B$\ddot{o}$ttcher
 \& Bloom 1998; Madejski et al. 1998). A 14-year period was found in the 
 optical light curve (Fan et al. 1998b). During 1995 January 24 - February 14,
  BL Lacertae showed a flux of 
 F($>$100 MeV) = ( 40 $\pm$12) $\times 10^{-8}$ photon cm$^{-2}$ s$^{-1}$
 with a photon spectral index $\alpha_{ph}$ = 2.2$\pm$0.3.  The upper 
limit on the 
 flux at higher energy is F($>$300 GeV) $<$ 0.53 $\times 10^{-11}$ 
 photon cm$^{-2}$ s$^{-1}$ ( Catanese et al. 1997). During the 1997
January 
 15/22 observation period, it was detected at a flux of  F($>$100 MeV) = 
 ( 171 $\pm$42) $\times 10^{-8}$ photon cm$^{-2}$ s$^{-1}$, with  a photon 
 spectral index $\alpha_{ph}$ = 1.68$\pm$0.16 and a dramatic
increase of a factor of 2.5  within a timescale of 8hrs, $\Delta T_{D}$ = 3.2 hrs.  
 Besides, simultaneous optical and $\gamma$-ray flares were
 observed ruling out external scattering models (see Bloom et al. 1997). 
 The observations from the object show that the spectrum of BL Lacertae
 hardens when the $\gamma$-ray flux increases. $f_{2KeV} = 0.749 \mu$ Jy 
 and $\alpha_{OX} = 1.31$ are reported in the
 paper by Perlman et al. (1996).

\section{Method and results}

 \subsection{Method}
 
Now we describe our method of estimating the basic parameters ($M$,
$\delta$, $\Phi$ and $d$) of the blazars with short timescale
variabilities in the $\gamma$-ray band. As mentioned above, high energy
$\gamma$-rays can escape only when the optical depth of $\gamma$-$\gamma$
 pair production is
not larger than
unity. Based on Becker \& Kafatos (1995), we can
obtain an approximate
relation for the optical depth at an arbitrary
angle, $\Phi$,
\begin{equation}
\tau_{\gamma\gamma}(M_{7},\Phi,d) = {\frac{1}{3}}(51-8\omega) \times
\Phi^{2.5}({\frac{d}{R_{g}}})^{-{\frac{2\alpha_{X}+3}{2}}}
+ kM_{7}^{-1}({\frac{d}{R_{g}}})^{-2\alpha_{X}-3}\;\;, 
\label{tau}
\end{equation}
where $k$ is given by
\begin{eqnarray}
k=&4.50\times10^{9}{\frac{\Psi(\alpha_{X})(2-\omega)(1+z)^{3+\alpha_{X}}F'_{0}(1+z-\sqrt{1+z})^{2}}{(2\alpha_{X}+4-\omega)(2\alpha_{X}+3)}} 
\times\nonumber\\
&[{\frac{({\frac{R_{0}}{R_{g}}})^{2\alpha_{X}+4-\omega}-({\frac{R_{ms}}{R_{g}}})^{2\alpha_{X}+4-\omega}}{({\frac{R_{0}}{R_{g}}})^{2-\omega}-({\frac{R_{ms}}{R_{g}}})^{2-\omega}}}]({\frac{E_{\gamma}}{4m_{e}c^{2}}})^{\alpha_{X}}\;\;,
\end{eqnarray}
$\Psi(\alpha_{X})$ is a function of the X-ray spectral index,
 $\alpha_{X}$, $F'_{0}$ the X-ray flux
parameter
in units of cm$^{-2}$~s$^{-1}$, $m_{e}$ the electron mass, $c$ the speed
of
light,
$R_{g}={\frac{GM}{c^{2}}}$ the Schwarzschild radius,
 $E_{\gamma}$ the average energy of the $\gamma$-rays, and $R_{0}$ and
$R_{ms}$ are,
respectively, the outer and inner radii of the accretion disk. $\omega$
 is a free parameter, $\omega = 3$ is for a two-temperature disk while
 $\omega = 0$ is for a uniformly bright disk.

From Eq. (1), the optical depth depends on $d$, $\Phi$ and 
$M$. At first, $d$ can be determined if the variability timescale
($\Delta T_{D}$) for a
blazar is observed, it is given by
\begin{equation}
{\frac{d}{R_{g}}} =1.73\times10^{3} {\frac{\Delta T_{D}}{1+z}}\delta
M_{7}^{-1}
\label{dis}
\end{equation}
Furthermore, using the observed $\gamma$-ray flux,
$F_{\gamma}^{obs}(>100MeV)$ in
units of ergs~cm$^{-2}$~s$^{-1}$, the relationship among 
 the intrinsic luminosity, $L_{in}$, the Doppler factor, $\delta$, the
mass of the central black hole, $M$,
 and the propagation angle, $\Phi$, is given by
$F_{\gamma}^{obs}(>100{\hbox{MeV}})=(1+z)^{1-\alpha_{\gamma}}\delta^{\alpha_{\gamma}+4}L_{in}/
\Omega D^2$. We can define an isotropic luminosity as 
$L_{iso}= 4\pi~D^{2}(1+z)^{\alpha_{\gamma}-1}F_{\gamma}^{obs}
(>100MeV)$ in units of $10^{48}$ ergs~s$^{-1}$, it can be expressed as
\begin{equation}
 L_{iso}^{48} =
{\frac{\lambda
2.52~10^{-3}\delta^{\alpha_{\gamma}+4}}{1-cos\Phi}}M_{7}\;\;,
\label{lumi}
\end{equation} 
where $L_{in} = \lambda L_{Edd}=\lambda 1.26\times
10^{45}M_{7}$, and $\lambda$ is a parameter depending on the specific
$\gamma$-ray emission model.

Substituting Eqs. (3) and (4) into Eq. (1), we obtain a function 
of $M$ and $\Phi$. From this
equation, 
a minimum value of $\tau_{\gamma\gamma}$ for a given mass, $M$, can be
determined by
${\frac{\partial\tau}{\partial\Phi}}|_{M} =0$, i.e. solving 
\begin{eqnarray}
{\frac{2.5}{3}}(51-8\omega)\Phi^{1.5}(1-cos\Phi) -
{\frac{1}{3}}(51-8\omega)\times{\frac{2\alpha_{X}+3}{2\alpha_{\gamma}+8}}\Phi^{2.5}sin\Phi&\nonumber
\\ 
-{\frac{2\alpha_{X}+3}{\alpha_{\gamma}+4}}kM_{7}^{-1}A^{-{\frac{2\alpha_{X}+3}{2}}}(1-cos\Phi)^{-{\frac{2\alpha_{X}+3}{2\alpha_{\gamma}+8}}}sin\Phi=
0 &
\label{partau}
\end{eqnarray}
where 
$$A = 1.73\times10^{3}{\frac{\Delta
T_{D}}{1+z}}M_{7}^{-{\frac{\alpha_{\gamma}+5}{\alpha_{\gamma}+4}}}({\frac{L_{iso}^{45}}{\lambda2.52}})^{{\frac{1}{4+\alpha_{\gamma}}}}$$

 Finally, letting the minimum of $\tau(M_{7},\Phi)$
 equal to 1.0, we have
 \begin{equation}
{\frac{1}{3}}(51-8\omega) \times
\Phi^{2.5}({\frac{d}{R_{g}}})^{-{\frac{2\alpha_{X}+3}{2}}} +
kM_{7}^{-1}({\frac{d}{R_{g}}})^{-2\alpha_{X}-3} = 1
\label{taueq1}
\end{equation}

 For a source with available data in the X-ray and $\gamma$-ray bands, the
 masses of the central black holes, $M_{7}$, the Doppler factor, $\delta$,
 the distance (height), $d$, and the propagation angle with respect to the
axis of
 the accretion disk, $\Phi$, can be derived from Eqs. (3),
(4), (5) and (6), where $R_{ms} = 
6R_{g}$, $R_{0}=30R_{g}$, $E_{\gamma}$= 1GeV and $\omega$ = 3 (a
two-temperature disk) are used.

\subsection{Results}

 Since we do not know the intrinsic $\gamma$-ray luminosity, we assume
 it is close to the Eddington luminosity, say $\lambda L_{Edd.}$. Using
the available X-ray and $\gamma$-ray data (see Table 1), we estimate
the four
parameters ($M_{7}$, $\Phi$, $\delta$, $d$) and find that the derived
values of the four parameters are not sensitive to the value of
$\lambda$.
The results are shown in Table 2.
In Table 1 Col. 1, gives the name;  
 Col. 2, the  redshift; Col. 3. the X-ray flux density in units of
$\mu$ Jy, Col. 4, the X-ray spectral index, $\alpha_{X}$. The averaged
 value of $<\alpha_{X}> = 0.67$ (Comastri et al. 1997) is adopted for
 PKS 1622-297, and we take $\alpha_{OX}=1.31$  for $\alpha_{X}$ for
 BL Lacertae, as did Ghisellini et al. (1998). 
Col. 5, the flux F($>$100MeV)
in units of 
 $10^{-6}$ photon cm$^{-2}$ s$^{-1}$,  $\sigma$ is the uncertainty; 
 Col. 6, $\gamma$-ray spectral index,  $\alpha_{\gamma}$ = 1.0
 is adopted for 0537-441 (see Fan et al. 1998c); Col. 7, the doubling time 
 scale in units of hours; Col. 8, the observed isotropic luminosity in
 units of $10^{48}$erg~s$^{-1}$.
 In Table 2, Col. 1 gives the name, Col. 2
 the Doppler factor ($\lambda$ = 1.0); Col. 3 the Doppler factor
($\lambda$ =
0.1); Col. 4, the central black hole mass in
units of 
 $10^{7}M_{\odot}$ ($\lambda$ = 1.0); Col. 5, the central black hole mass
 ($\lambda = 0.1$); Col. 6, the propagation angle, $\Phi$ in the units of
 degree($^{\circ}$) ($\lambda = 1.0$),Col. 7, the propagation angle
($\lambda = 0.1$), Col. 8, the distance (height), ${\frac{d}{R_{g}}}$,
where the $\gamma$-rays are created ($\lambda = 1.0$); Col. 9, the
distance (height) ($\lambda = 0.1$); Col. 10, the
central black hole mass estimated by the method of
 Dermer \& Gehrels (1995), in units of $10^{7}M_{\odot}$; 
 Col. 11, the mass  estimated directly from the Eddington limit in 
 units of  $10^{10}M_{\odot}$ (cf. section 4.1).

 For comparison, we also considered a  uniformly-bright disk for 3C279.
 The derived basic parameters are listed in Table 3, in which
 Col. 1 gives the flare
 time, Col. 2 the parameter  $\lambda$, Col. 3 the Doppler factor,
 Col. 4 the masses of the central black hole, 5 the propagation angle, and
 Col. 6 the distance, $\frac{d}{R_{g}}$. There is not much difference
 between the parameters obtained for a two-temperature disk ($\omega = 3$) and those for
 a uniformly-bright disk ($\omega = 0$) although the masses are a little
 smaller, the propagation angles are wider, and the distance is farther
 as compared with those obtained for $\omega = 3$.

 In the paper by Becker \& Kafatos (1995), basic parameters are also
 determined for the 1991 flare of 3C279. Using $M=10^{9}M_{\odot}$,
 ${\frac{R_{0}}{R_{g}}}$ = 30, ${\frac{R_{ms}}{R_{g}}}$ = 6, and
 $\omega$ = 3, Becker \& Kafatos obtained a variability time scale of
 $t_{\gamma\gamma}$ = 2.57 $days$, ${\frac{d}{R_{g}}}$ = 45.2 and
 $\Delta \phi = 9.5^{\circ}$.  But from our definition of the variability
 time scale, $t_{\gamma\gamma}$
 should be 0.5 days for the 1991 flare. In this sense, our results should
 correspond to the first case presented in the first line in Table 1
 ( Becker \& Kafatos 1995 ).  From  the 3C279 1991 flare, the masses
obtained
 here are $ (0.667 \sim 1.062) \times 10^{8} M_{\odot}$, which are consistent
 with the value of $10^{8}M_{\odot}$ (Becker \& Kafatos 1995), but our
 results for the distance, $\frac{d}{R_{g}}$, and the propagation angle, $\Phi$,
 are larger than theirs. These differences do not suggest that our results
 are not consistent with theirs because the methods used are different.

\begin{table*}
\caption{ Observation data for $\gamma$-ray loud blazars}
\begin{tabular}{lcccccccccccc} 
\hline\noalign{\smallskip}
 $Name$ & $z$ & $f_{1KeV}$&Ref& $\alpha_{X}$&Ref& $ F(\sigma)$ &Ref&
$\alpha_{\gamma}$
&Ref&
$\Delta T_{D}$ &Ref& $L_{iso}^{48}$ \\
(1)     & (2)  & (3)  & (4) & (5) & (6) & (7) & (8)&
(9)&(10)&(11)&(12)&(13) \\
\noalign{\smallskip} \hline
0208-512  & 1.003& 0.61 & C97& 1.04(0.04) &C97
& 9.1(0.4)&B93   & 0.69 &B93& 5.6 &S96 & 2.0  \\

0528+134 & 2.07  & 0.65 &C97& 0.54(0.29)&C97 
&3.08(0.35) &W97 & 1.21 &W97& 24. &W97 & 18.4  \\ 

0537-441 & 0.894 & 0.81 &C97& 1.16(0.09) &C97
& 2.0(0.4) &H96  & 1.0 &F98 & 16. &H96 & 3.01 \\

1253-055 & 0.537 & 2.43 &H96b& 0.68 &H96b       
& 2.8(0.4) &K93  & 1.02 &K93& 12. &K93 & 1.34 \\

1253-055 & 0.538 & 2.0 &L98 & 0.78 &L98      
& 11.(1.) &M96   & 0.97&M96 & 6. &W98  & 5.75 \\

1622-297 & 0.815 & 0.08 &M97& 0.67 &C97
& 17.(3.) &M97   & 0.87 &M97& 4.85 &M97& 26.9 \\

1633+382 & 1.814 & 0.42&C97 & 0.53(0.08)&C97 
&0.96(0.08)&M93  & 0.86 &M93& 16. &M93 & 9.72  \\

2200+420 & 0.07  & 1.84 &P96& 1.31 &P96      
& 1.71(0.42)&B97 & 0.68 &B97& 3.2 &B97 & 0.019 \\
\hline 
\end{tabular}\\
 Notes to Table 1\\
 Col. 1, gives the name; Col. 2, the  redshift; Col. 3. the X-ray flux
 density in units of $\mu$ Jy; Col. 4, reference for Col. 3; 
 Col. 5. the X-ray spectral index,
 $\alpha_{X}$. The averaged value of $<\alpha_{X}> = 0.67$
 (Comastri et al. 1997) is adopted to PKS 1622-297, and
 $\alpha_{OX}=1.31$ is used for $\alpha_{X}$ for BL Lacertae as
 did Ghisellini et al. (1998); Col. 6, references for Col. 5;
 Col. 7, the flux F($>$100MeV) in units of 
 $10^{-6}$ photon cm$^{-2}$ s$^{-1}$,  $\sigma$ is the uncertainty; 
 Col. 8, the reference for Col. 7;
 Col. 9, $\gamma$-ray spectral index,  $\alpha_{\gamma}$ = 1.0
 is adopted for 0537-441 (see Fan et al. 1998c); 
 Col. 10, reference for Col. 9;
 Col. 11, the doubling time  scale in units of hours; 
 Col. 12, references for Col. 11;
 Col. 13, the observed isotropic luminosity in
 units of $10^{48}$ ergs s$^{-1}$ calculated in the present paper.\\
 B93: Bertsch et al. 1993; B97: Bloom et al. 1997; 
 C97: Comastri et al. 1997; F98: Fan et al. 1998c;
 H96: Hartman 1996; H96b: Hartman et al. 1996; 
 K93: Kniffen et al. 1993; L98: Lawson \& McHardy 1998;
 M93: Mattox et al. 1993; M96: McHardy 1996; M97: Mattox et al. 1997;
 P96: Perlman et al. 1996; S96: Stacy et al. 1996;
 W97: Wagner et al. 1997;
 W98: Wehrle et al. 1998.
\end{table*}

\begin{table*}
\caption{ Determined Results for 7 $\gamma$-ray loud Blazars }
\begin{tabular}{lcccccccccc} 
\hline\noalign{\smallskip}
 $Name$ & $\delta$ & $\delta$  &  $M_{7}$ & $M_{7}$ &$\Phi$& $\Phi$ &
${\frac{d}{R_{g}}}$&${\frac{d}{R_{g}}}$ &
$M_{7}^{KN}$ & $M_{10}^{T}$ \\ 
  & $(\lambda=1.0)$ & $(\lambda=0.1)$ &$(\lambda=1.0)$ & $(\lambda=0.1)$
&$(\lambda=1.0)$ & $(\lambda=0.1)$ &$(\lambda=1.0)$ & $(\lambda=0.1)$ & &
\\  (1)     & (2)  & (3)  & (4) & (5) & (6) & (7) & (8) & (9) & (10)
&(11)\\
\noalign{\smallskip} \hline
0208-512 & 1.00 & 1.33 & 82.94 & 131.5 & 25.5 & 21.7 & 71.9 & 61.3 & 14.4
&
1.59\\
0528+134 & 3.72  & 5.33 & 5.09  & 8.21  & 43.  & 39.2 & 411 & 366  &
22.98&
14.6
\\ 
0537-441 & 1.83 & 2.51 & 12.56 & 19.02 & 38.1 & 35.3&  86.7 & 80.3  & 3.48
&
2.39\\
1253-055 & 1.43 & 2.03 & 6.67  & 10.62 & 21.3 & 19.2&  121. & 107.  & 1.92
&
1.06\\
1253-055 & 2.11 & 3.00 & 5.40  & 8.47  & 23.8 & 21.7 & 110. & 99.6& 7.53 &
4.56\\
1622-297 & 2.42 & 3.45 & 5.71  & 9.06  & 14.8 & 13.5&  81. & 73. & 25.  &
21.35\\
1633+382 & 3.22 & 4.60 & 3.81  & 6.15  & 36.8 & 33.5 & 347. & 309. & 5.74
& 7.71
\\
2200+420 & 0.57 & 0.79 & 4.45  & 6.63  & 14.1 & 13.0 & 27.7 & 25.8 & 0.02
&
0.015\\
\hline 
\end{tabular}\\
 Notes to Table 2:\\
 Col. 1 gives the name, Col. 2 Doppler factor ($\lambda$ = 1.0);
 Col. 3 Doppler factor ($\lambda$ = 0.1);
 Col. 4, the central black hole mass in units of 
 $10^{7}M_{\odot}$ ($\lambda$ = 1.0); Col. 5, the central black hole mass
 ($\lambda = 0.1$); Col. 6, propagation angle, $\Phi$ in the units of
 degree($^{\circ}$) ($\lambda = 1.0$),Col. 7, propagation angle
($\lambda = 0.1$), Col. 8, the distance (height), ${\frac{d}{R_{g}}}$,
 where the $\gamma$-rays are created ($\lambda = 1.0$);
 Col. 9, the distance (height) ($\lambda = 0.1$);
 Col. 10, the central black hole mass estimated from the method of
 Dermer \& Gehrels (1995), in units of $10^{7}M_{\odot}$; 
 Col. 11, the mass  estimated directly from Eddington limit in 
 units of  $10^{10}M_{\odot}$ (cf. section 4.1).
\end{table*}

\begin{table*}
\caption{ Determined Results for 3C 279 (1253-055) for a uniformlly-bright disk }
\begin{tabular}{lccccc} 
\hline\noalign{\smallskip}
 $Flare$ & $\lambda$ & $\delta$  &  $M_{7}$ & $\Phi$ & ${\frac{d}{R_{g}}}$\\
  (1) & (2) & (3) & (4)  & (5) & (6) \\ \hline
 1991 & 0.1 & 2.1 & 7.64 & 20. & 154.\\
 1991 & 1.0 & 1.5 & 4.93 & 21.8 & 171.7 \\
 1996 & 0.1 & 3.14 & 6.19 & 23.15 & 143\\
 1996 & 1.0 & 2.23 & 4.0 &  25.2 & 157 \\
\hline 
\end{tabular}\\
 Notes to Table 3\\
 Col. 1 gives the flare time, Col. 2 the parameter of $\lambda$,
 Col. 3 the Doppler factor, Col. 4 the masses of the central black holes,
 Col. 5 the propagation angle, and Col. 6 the distance, $\frac{d}{R_{g}}$.
\end{table*}

 \section{Discussion}

 Rapid $\gamma$-ray variability has been detected in 10 $\gamma$-ray
 loud objects including 2 known TeV emitting objects, but we  only discuss
 the absorption effect of X-rays on the GeV $\gamma$-ray emissions. 
 Neither TeV
 emitting object is  included in the present paper since there is no
 evidence of GeV $\gamma$-ray variability for them even though 
 rapid TeV variabilities have been detected from them
  ( Gaidos et al. 1996; Quinn et al. 1996).  PKS 1406-074 shows rapid GeV
$\gamma$-ray variability (Wagner et al. 1995), but there is
 no X-ray data available in the literature. So, only seven objects are 
 considered here. It is important to note that our model results are 
 independent of  the emission mechanisms of $\gamma$-rays although they
 are dependent on the X-ray emission mechanism from the accretion disk.
 For X-ray emission, many authors have discussed the processes. Apart
 from the above mentioned process, the synchrotron/inverse Compton process 
 (see 
 Konigl 1981; Kubo et al. 1998;
 Padovani et al. 1997; Pian et al. 1998; 
 Makino et al. 1997;
 Sambruna et al. 1996;  Urry et al. 1986) and the annihilation of
 $e^{\pm}$ (Robson 1986) are also important mechanisms of X-ray
 production.
 In addition, fluorescent and  thermal bremsstrahlung emissions
 are important for X-rays (Rybicki \& Lightman 1979; Makino 1999).
 
 \subsection{Mass}

 Assuming that the  observed  $\gamma$-ray luminosity is isotropic without
 a beaming effect and equals the 
 Eddington-luminosity, one can estimate the central black hole mass,
 \begin{equation}
 M_{10} \geq {\frac{L_{T}}{1.26\times 10^{48} erg s^{-1}}}\;\;,
 \end{equation} 
 where L$_T$ is the bolometric luminosity for emission in the Thomson region
 and
 M$_{10}$ is the central black hole mass in units of 10$^{10}M_{\odot}$. 
 The derived masses are as high as 10$^{11}M_{\odot}$ for some $\gamma$-ray 
 loud blazars, PKS 0528+134 and PKS1622-297 for instance (see Col. 11
 in Table 2). 

  Dermer \& Gehrels (1995) considered the Klein-Nishina effect without
 the beaming effect and  obtained an expression for the  black hole mass,

 \begin{equation}
 M_{8}^{KN} \geq {\frac{3\pi d_{L}^{2}(m_{e}c^{2})}{2 \times 1.26\times 
 10^{46} erg s^{-1}}}{\frac{F(\varepsilon_{l},\varepsilon_{u})}{1+z}}
 ln[2\varepsilon_{l}(1+z)]
 \end{equation} 
 where $F(\varepsilon_{l},\varepsilon_{u})$ is the integrated photon flux 
 in units of 10$^{-6}$ photon cm$^{-2}$ s$^{-1}$ between photon energies  
 $\varepsilon_{l}$ and $\varepsilon_{u}$ (in units of 0.511MeV). 
 For the objects considered here,  $M_{7}^{KN}$ is obtained and 
 shown in Col. 10 in table 2.  Table 2 shows that there are some
differences between the masses obtained from our method and those
estimated from the method of  Dermer \& Gehrels (1995) for PKS 0208-512,
PKS0528+134, PKS 1622-297 and BL Lacertae. The reason is that they 
 considered the Klein-Nishina effect but not the beaming
 effect nor the actual solid angle. In addition, other reasons can also
 cause the difference. For PKS1622-297, if we adopt a
 flux density of $2.45 \times 10^{-6}$ photon cm$^{-2}$ s$^{-1}$
 instead of the peak value as did Muhkerjee et al. (1997) 
 and Fan et al. (1998c), then the isotropic luminosity is $3.87\times 10^{48}$ 
 erg s$^{-1}$ suggesting a mass of 6.1$M_{7}$ from our calculation and a 
 mass of 3.61$M_{7}$ from the method of Dermer \& Gehrels (1995).
 Both masses are quite similar in this  case.  For 3C279, our results 
 show that the estimated central black hole masses, 6.67$M_{7}$ and 5.40
$M_{7}$ if $\lambda = 1 $ is adopted, and 10.62$M_{7}$ and 8.47$M_{7}$ 
if $\lambda = 0.1$  is adopted, for the 1991 and 1996 flares
respectively, are almost  the same. The difference in masses obtained 
from our method when $\lambda$ changes by an order of one magnitude is  
only a factor of $\sim$1.5, suggesting that the estimated mass is not 
 sensitive to the choice of $\lambda$. The masses obtained here are in a
range of ( $\sim$4 -  83)$\times 10^{7}M_{\odot}$ ($\lambda = 1.0$) 
or ($\sim$6 - 130)$\times 10^{7}M_{\odot}$ ($\lambda = 0.1$).

 For 3C279,  to fit the multiwavelength energy spectrum corresponding 
 to the 1991 $\gamma$-ray flare, Hartman et al. (1996) used an accreting
black
 hole of $10^{8}M_{\odot}$ while  Becker \& Kafatos (1995) also obtained
 $M = 10^{8-9} M_{\odot}$; our result of  $ M = (6.67-10.62)\times
 10^{7}M_{\odot}$ is consistent with theirs. For PKS 0537-441, Romero et
 al. (2000) obtained a mass of $8\times 10^{7}M_{\odot}$ for the
 central black hole, our result of $12.56 \times 10^{7}M_{\odot}$
 ($\lambda= 1.0$) does not conflict with theirs, the slight difference
 is from that fact that we used an outer radius of a two-temperature
 disk  $R_{0} = 30R_{g}$ while they adopted
 $R_{0}=100R_{g}$.

\subsection{ Beaming factors}

 To explain the extremely high and violently variable luminosity of AGNs, 
 a beaming model has been proposed. 
 To let the optical depth ($\tau_{\gamma\gamma}$)  be less than unity,
 a Doppler factor in the $\gamma$-ray region has been obtained for some
objects 
 by other authors.
  $\delta \geq 7.6$ for Q1633+382 ( Mattox et al., 1993);  
 $\delta \geq 6.3 - 8.5$
 for the 3C279 1996 flare ( Wehrle et al. 1998) and  $\delta \geq 3.9 $
for 
 the 3C279 1991 flare ( Mattox et al. 1993), $\delta \sim 5 $ is also 
 obtained by Henri et al. (1993); 
 $\delta \geq 6.6 \sim 8.1$ for
 PKS 1622-297 (Mattox et al. 1997).  The Doppler factors obtained here
 are smaller than those obtained by others. The reason is that we believe
 that the $\gamma$-rays are from a solid angle $\Omega = 2\pi(1-cos\Phi)$ 
 while others assume that the $\gamma$-rays are isotropic, i.e.,
 $\Omega = 4\pi$. 
 Therefore, different assumptions result in different Doppler factors.
 The Doppler factor derived under the former assumption is 
 $(\frac{1-cos\Phi}{2})^{1/(4+\alpha)}$
 times the factor derived under the latter assumption.

\subsection{Propagation angle, $\Phi$}
 
 Generally, the observed luminosity is calculated assuming the emissions
 are isotropic. From the arguments ( Becker \& Kafatos, 1995 and
reference therein ), we know that only the $\gamma$-rays within the
 propagation angle are visible, i.e., $\tau_{\gamma\gamma} \leq 1.0$.
 Our calculations show that the $\Phi$'s are in the range of
 14$^{\circ}$ to 43$^{\circ}$ ($\lambda = 1.0$) or 13$^{\circ}$ to
39$^{\circ}$ ($\lambda = 0.1$); the average value 
 is $<\Phi> = 27^{\circ}.5$ ($\lambda = 1.0$) or $<\Phi> = 24^{\circ}.6$
($\lambda = 0.1$) . These values of $\Phi$ are consistent with the X-ray 
 cone ($\Phi = 15^{\circ}-40^{\circ}$) of BL Lac objects ( Maraschi \& 
 Rovetti, 1994).  
 
\subsection{Distance, ${\frac{d}{R_{g}}}$}

 The variability time scale places some constraints on the size of the 
 emission region. By fitting the energy spectrum of 3C279, Hartman et al.
 (1996) obtained that the $\gamma$-rays in 3C279 were from 100$R_{g}$. 
 From the time scale and the estimated black hole mass of Mkn421, Xie et
 al. (1998b) found that the $\gamma$-rays in Mkn421 are from 205$R_{g}$. 
 Recently, Celotti \& Ghisellini (1998) argued that the $\gamma$-rays are
 from a distance of hundreds of Schwarzschild radii. From our method,
 the distance is from 25$\sim$28$R_{g}$ for BL Lacertae to
366$\sim$411$R_{g}$ for
 PKS0528+134. The average distance is 160$R_{g}$ ($\lambda = 1.0$) or 143
 $R_{g}$ ($\lambda = 0.1$) for the seven objects.   

 For the X-ray emission regions, the simultaneous variability of X-rays
 and $\gamma$-rays during 1996 flare of 3C 279 suggests that the time
scales
 can be taken the same for both bands. Then the size of the X-ray
 emission region should not be much smaller than the size of $\gamma$-ray
 emission region. For BL Lacertae, there is an X-ray variation up to 30\% on a time
 scale of hours (Kawai et al. 1991) suggesting that the size of the X-ray
 emission region is $130 R_{g}$ if we take
 $M = (4.45 \sim 6.63) \times 10^{7} M_{\odot}$, which is much
 larger than the size of the $\gamma$-ray emission region. For two other
 objects (PKS 0528+134 and PKS 0537-441), variations in the X-ray
 region are detected but there is no  variability 
 timescale of days reported
 in the literature (see Ghisellini et al. 1999; Treves et al. 1993)
 except for a variation of 50\% in 2 weeks for 0528+134
 (Ghisellini et al. 1999). For the remaining three objects, there are
 no X-ray timescales. Anyway,  the available  X-ray variability 
 timescales of the three objects ( 3C279, BL Lacerte, and 0528+134)
 imply that the size of the
 X-ray emission region is not much smaller than the size of 
 the $\gamma$-ray emission region.

\subsection{Summary}
 In this paper, the optical depth of a $\gamma$-ray travelling in
 the field of a two-temperature disk including the beaming effect have
been used
 to determine the central mass, $M$, Doppler factor, $\delta$,
 propagation angle, $\Phi$, and the distance, $d$  for seven
 $\gamma$-ray loud blazars with available short $\gamma$-ray timescales.
 The masses obtained  are compared with
 those obtained by the method of Dermer \& Gehrels (1995).
 The distance (height) on average is 140 $\sim$ 160 $R_{g}$ from the
 center for the seven objects,  the propagation angle $\Phi$ is, on 
average,
 24$^{\circ}$$\sim$27$^{\circ}$.  For 3C279, a uniformly-bright disk
 is also considered to determine the basic parameters, the results are
 compared with those by Becker \& Kafatos (1995), the masses obtained
 from the two flares being almost the same.

\section*{Acknowledgements} 
 We thank Prof. S. Y. Pei for the useful discussion,
 Prof. P.K. MacKeown for his critical reading of the manuscript, Drs. P.
Becker,
 M. B$\ddot{o}$ttcher, P.M. Chadwick, C. Dermer, G. Ghisellini, F. Makino, 
 and S. Wagner for  sending us 
 their publications and information, and an anonymous referee
 for many useful comments and suggestions.
  This work is supported by 
 an RGC grant of Hong Kong, the 
 National Pan Deng Project of  China and the National Natural
 Scientific Foundation of China.

\end{document}